\documentclass[conference]{IEEEtran}
\usepackage{cite}
\usepackage[pdftex]{graphicx}
\usepackage[cmex10]{amsmath}
\usepackage{array}
\usepackage{subcaption}
\usepackage[utf8]{inputenc}
\DeclareUnicodeCharacter{A0}{~}
\usepackage[T1]{fontenc}
\usepackage{fixltx2e}
\usepackage{url}
\usepackage{hyperref}
\hypersetup{colorlinks=true, linkcolor=black, citecolor=black, filecolor=black, urlcolor=black, bookmarksdepth=subsection}
\usepackage{booktabs}
\usepackage[autostyle,english=british]{csquotes}
\usepackage{booktabs}
\usepackage{docmute}
\usepackage{varioref}
\usepackage{mathtools}
\usepackage{algorithm}
\usepackage[noend]{algpseudocode}

\title{PoliFi: Airtime Policy Enforcement for WiFi}

\author{\IEEEauthorblockN{Toke Høiland-Jørgensen} \IEEEauthorblockA{Dept. of
    Computer Science\\ Karlstad University, Sweden\\
    toke.hoiland-jorgensen@kau.se}
  \and
  \IEEEauthorblockN{Per Hurtig} \IEEEauthorblockA{Dept. of
    Computer Science\\ Karlstad University, Sweden\\
    per.hurtig@kau.se}
  \and
  \IEEEauthorblockN{Anna Brunstrom} \IEEEauthorblockA{Dept. of
    Computer Science\\ Karlstad University, Sweden\\
    anna.brunstrom@kau.se}}
\date{\today}

\begin{document}

\bstctlcite{myctl}
\maketitle

\begin{abstract}
  As WiFi grows ever more popular, airtime contention becomes an increasing
  problem. One way to alleviate this is through network policy enforcement.
  Unfortunately, WiFi lacks protocol support for configuring policies for its
  usage, and since network-wide coordination cannot generally be ensured,
  enforcing policy is challenging.

  However, as we have shown in previous work, an access point can influence the
  behaviour of connected devices by changing its scheduling of transmission
  opportunities, which can be used to achieve airtime fairness. In this work, we
  show that this mechanism can be extended to successfully enforce airtime usage
  policies in WiFi networks. We implement this as an extension our previous
  airtime fairness work, and present PoliFi, the resulting policy enforcement
  system.

  Our evaluation shows that PoliFi makes it possible to express a range of
  useful policies. These include prioritisation of specific devices; balancing
  groups of devices for sharing between different logical networks or network
  slices; and limiting groups of devices to implement guest networks or other
  low-priority services. We also show how these can be used to improve the
  performance of a real-world DASH video streaming application.
\end{abstract}


\section{Introduction}\label{sec:introduction}
WiFi is increasingly becoming the ubiquitous connectivity technology in homes as
well as in enterprises. The ability for anyone to set up an access point and
connect any device to it is one of the driving factors behind this increase of
popularity. However, increased popularity also means increased contention for
resources as more devices are deployed.

Since no two devices can transmit at the same time on a given frequency, the
sparse resource that determines performance in WiFi networks is the time spent
transmitting, also known as airtime usage. The 802.11 Media Access Control (MAC)
protocol used in WiFi networks does not, in itself, guarantee a fair usage of
this sparse resource. In fact it is well known that devices transmitting at
lower rates can use more than their fair share of the
airtime~\cite{heusse_performance_2003}.

One way to improve performance of a network under contention is to apply
different policies to different devices on the network, which works best if
applied directly to the sparse resource instead of a proxy such as byte-level
throughput. However, WiFi is decentralised at the protocol level, and thus lacks
protocol support for enforcing policies on airtime usage. Fortunately, it turns
out that in the common infrastructure deployment scenario, the access point can
exert quite a bit of influence on the transmission behaviour of clients, or
\emph{stations}, as they are commonly called. In previous work, we have shown
that this makes it possible to achieve airtime fairness between stations in a
WiFi network by making appropriate scheduling decisions at the
AP~\cite{usenix-paper}. Given such a mechanism to enforce fairness, a natural
question is whether it can be extended to express different capacity sharing
policies. In this work we answer this question in the affirmative, in the form
of a workable solution to airtime policy enforcement in WiFi, which we have
named \emph{PoliFi}.

The number of possible policies one might want to express is all but infinite.
Therefore, to focus our discussion, we define the following three representative
policy use cases:

\begin{enumerate}
\item Prioritising devices. It should be possible to configure one or more
  devices to receive a higher share of network resources than other devices on
  the network.
\item Balancing device groups. In this use case, the network should be
  configured to share the available resources between groups of devices in a
  given way. For instance, this could be used to implement the ``network
  slicing'' concept often seen in 5G architectures~\cite{foukas2017network}.
\item Limiting groups of devices to a \emph{maximum} capacity share, in order to
  implement a lower-priority service, such as a guest network.
\end{enumerate}

PoliFi makes it possible for the user to express all of these policies. Our
design builds on our previous airtime scheduler for the Linux kernel, but
extends it by (a) generalising the implementation from a specific driver to the
common kernel WiFi stack, (b) extending the kernel scheduler to support weighted
scheduling of stations, and (c) adding a userspace policy daemon that transforms
the higher-level policy decisions into configuration of the kernel scheduling
mechanism.

The rest of this paper presents PoliFi in detail, and is structured as follows:
Section~\ref{sec:related-work} summarises related work.
Section~\ref{sec:policy-daemon} describes our design, with a performance
analysis presented in Section~\ref{sec:evaluation}. Finally,
Section~\ref{sec:conclusion} concludes.

\section{Related work}
\label{sec:related-work}
Network policies are, in general, nothing new. For instance, standardisation of
different traffic classes has occurred in the form of the DiffServ
framework~\cite{rfc4594}. In the WiFi world, the 802.11e standard defines
different priority levels, which can be mapped to DiffServ code
points~\cite{rfc8325}. However, this is all related to applying policies to
different types of traffic, whereas PoliFi deals with realising different
capacity sharing policies between devices on the same network at the airtime
usage level. As such, PoliFi is orthogonal so DiffServ, 802.11e and other
traffic class policy mechanisms.

As mentioned above, PoliFi is an extension of our previous work implementing an
airtime fairness enforcement mechanism in Linux~\cite{usenix-paper}. Compared to
this previous work, PoliFi adds the policy enforcement component, and also
generalises the mechanism by moving it out of the device drivers and into the
common WiFi subsystem in Linux, thus making it applicable to more device
drivers.

The concept of airtime policy enforcement appears in the concept of
\emph{network slicing}, which is an important part of the upcoming 5G mobile
network architecture~\cite{foukas2017network}. Network slicing involves
splitting up a network into several virtual parts that are conceptually isolated
from one another, which is a form of policy enforcement. A description of how to
achieve network slicing in WiFi networks is given in~\cite{wifi-slicing}, which
corresponds roughly to our second use-case. The authors implement a prototype in
simulation. Our mechanism builds on the same basic concept of computing
per-device weights from group weights, but we solve a number of issues that
prevent it from being implemented on real hardware. In addition,
\cite{wifi-slicing} only covers the second of our three policy use-cases.

Another approach to splitting a wireless network into multiple parts is
presented in~\cite{katsalis_virtual_2015}, which describes a scheme where a
separate software router is installed in the access point. This software router
queues packets and enforces capacity sharing. However, the capacity sharing is
implemented at the bandwidth level which, as mentioned above, is not the sparse
resource in a WiFi network.

A description of a scheme for network slicing in a home network is described
in~\cite{Yiakoumis:2011}. The authors describe a design that uses Software
Defined Networking (SDN) to split a home network into different parts, but do
not discuss any mechanism for how the sharing is achieved.

Finally, some enterprise APs offer features related to airtime fairness and
policy configuration, e.g., \cite{cisco-atf}. Unfortunately, no technical
description of how these policies are enforced is generally available, which
prevents us from comparing them to our solution.

\section{The PoliFi Design}
\label{sec:policy-daemon}
\begin{figure}[tb]
\centering
\includegraphics[width=\linewidth]{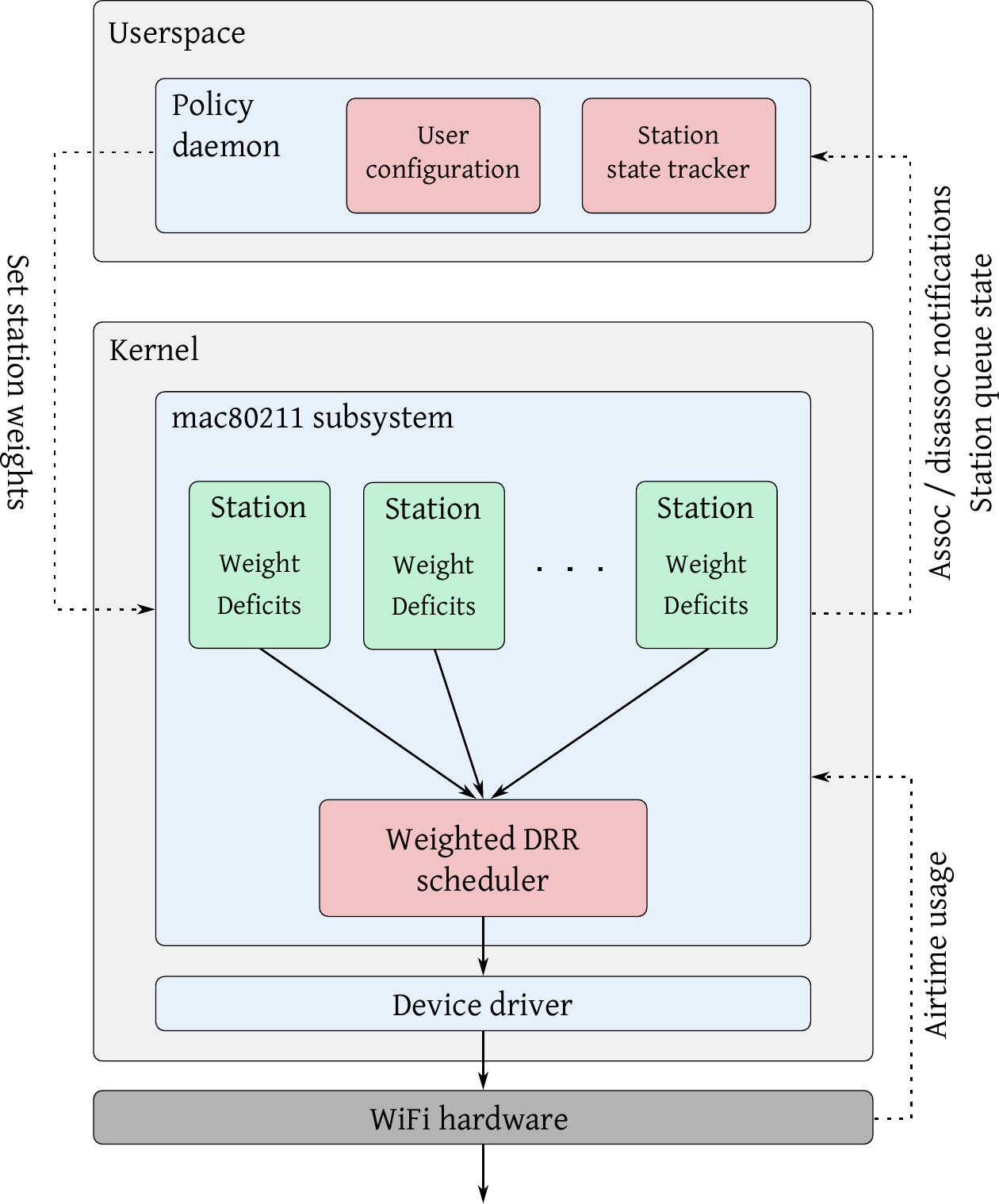}
\caption{\label{fig:policy-daemon-diagram} The high-level design of PoliFi. The
  kernel maintains data structures for every station, containing its current
  airtime deficits and configured weight. The scheduler uses this to decide
  which station to transmit to next. The hardware reports airtime usage on TX
  completion. The userspace daemon tracks the associated stations and their
  queue state, and updates the weights in the kernel based on user policy
  preferences.}
\end{figure}

We have designed PoliFi as a two-part solution, where a user-space daemon is
configured by the user, and in turn configures a scheduling mechanism in the
kernel. In this section, we describe our design in detail. A diagram of the
design is shown in Figure~\ref{fig:policy-daemon-diagram}. We begin by
describing the user space daemon that configures the policies. Following this,
we describe how the weighted Deficit Round-Robin (DRR) scheduling mechanism is
used to achieve the desired policies, and finally we describe how the mechanism
is integrated into the Linux kernel WiFi stack.

\subsection{Userspace Policy Daemon}
\label{sec:user-daemon}

We implement the userspace policy daemon as part of the \emph{hostapd} access
point management daemon. This is the daemon responsible for configuring wireless
devices in access point mode in Linux. This means it already implements policies
for other aspects of client behaviour (such as authentication), which makes it a
natural place to implement airtime policy as well.

The module we have added to hostapd can be configured in three modes,
corresponding to the three use cases described in the introduction: static mode,
dynamic mode and limit mode. The user can configure each of these modes per
physical WiFi domain, and assign parameters for individual stations (based on
their MAC addresses), or for entire Basic Service Sets (BSSes). The latter is a
natural grouping mechanism, since this corresponds to logical networks
configured on the same device (e.g., a primary and a guest network). However,
extending the design to any other logical grouping mechanism is straight
forward.

In static mode, the daemon will simply assign static weights to stations when
they associate to the access point. Weights can be configured for individual
stations, while a default weight can be set for each BSS, which will be applied
to all stations that do not have an explicit value configured. This implements
the basic use case of assigning higher priorities to specific devices, but does
not guarantee any specific total share.

The dynamic and limit modes work by assigning weights to each BSS, which are
interpreted as their relative shares of the total airtime, regardless of how
many stations they each have associated. Additionally, in limit mode, one or
more BSSes can be marked as \emph{limited}. BBSes that are marked as limited are
not allowed to exceed their configured share, whereas no limitations are imposed
on unmarked BSSes. Thus, dynamic mode implements the second use case, while
limit mode implements the third.

For both dynamic mode and limit mode, the daemon periodically polls the kernel
to discover which stations are currently active, using the queue backlog as a
measure of activity, as discussed below. After each polling event, per-station
weights are computed based on the number of active stations in each BSS, and
these weights are configured in the kernel. The details of the weight
computation, and how this is used to achieve the desired policy, is discussed in
the next section. Selecting the polling frequency is a tradeoff between reaction
time and system load overhead. The polling interval defaults to 100\,ms, which
we have found to offer good reaction times (see
Section~\ref{sec:dynamic-measurements}), while having a negligible overhead on
our test system.

While our implementation is focused on the single access point case, where the
access point enforces a single configured policy, the userspace daemon could
just as well pull its policy configuration from a central cloud-based management
service, while retaining the same policy enforcement mechanism.

\subsection{Weighted Airtime DRR}
\label{sec:weighted-drr}
The fairness mechanism that we are starting from (described in detail
in~\cite{usenix-paper}) is a Deficit Round-Robin scheduler, which operates by
accounting airtime usage as reported by the WiFi hardware after a transmission
has completed, and scheduling transmissions to ensure that the aggregate usage
over time is the same for all active stations. Using the airtime information
provided after transmission completes means that retransmissions can be
accounted for, which improves accuracy especially for stations with low signal
quality. Furthermore, as we have shown in our previous work, for TCP traffic we
can provide fairness even for transmissions transfers coming \emph{from} each
station. This is achieved by accounting the airtime of received packets, which
causes the scheduler to throttle the rate of TCP ACKs going back to the station.

\subsubsection{Adding Weighted Scheduling}
\label{sec:airtime-priority}

Given this effective airtime fairness scheduler, we can realise arbitrary
division of the available capacity between different stations, by simply
assigning them different scheduling weights. For the DRR scheduling algorithm
employed by our scheduler, this is achieved by using different quantums per
station. Thus, to apply this to airtime policy enforcement, we need to express
the desired policy as a number of different service weights for each of the
active devices.

The first use case is trivially expressed in terms of weights: simply assign the
prioritised device a higher weight; for instance, to double its priority, assign
it a weight of 2. The second use case, where capacity should be split between
groups of devices has been covered in the network slicing use case described
in~\cite{wifi-slicing}: each group is assigned a weight signifying its share
relative to the other groups; from these group weights, each device in that
group is assigned a weight computed by dividing the group weight with the number
of active devices in that group.

The final use case requires limiting one or more groups of stations to a fixed
share of the available capacity. This can be illustrated with the guest network
use case, where an example policy could be that a guest network is not allowed
to exceed 50\% of the available capacity. If this policy is implemented as a
fixed share between groups, however, a single station on the guest network would
be able to get the same capacity as, say, five users of the regular network,
which is not what we want. Thus, a different policy is needed: a group can be
limited, and should have its weight adjusted only if it would get \emph{more}
than the configured share, not if it gets less. Thus, this becomes a two-step
procedure that first assigns unit weights to all devices (which is the default
when no policy is applied), and calculates whether or not this results in the
limited group using more than its configured share of the airtime. If it does, a
policy is computed in the same way as for the dynamic use case, which results in
the limited group being assigned exactly its configured airtime share.

\subsubsection{Computing the Weights}
\label{sec:diff-assign-weights}
Having established that our desired policies can be expressed in terms of
weights, we turn to the practical difficulties of applying this to a real WiFi
system.

First, the approach outlined above assumes that we have knowledge of which
stations are active at any given time. This might look trivial at first glance,
since an access point needs to maintain some amount of state for all currently
associated clients in any case. However, clients can be associated to an access
point without sending or receiving any data, and thus without consuming any
airtime. This means that association state in itself is not sufficient to
ascertain the set of currently active clients. Fortunately, we have another
piece of data: The queue backlog for each device. Monitoring the backlog gives
us a straight-forward indicator for activity without having to monitor actual
packet flows; we can simply consider any device that has had a non-zero queue
backlog within a suitably short time span as active, and use that number in our
calculations.

The second difficulty lies in the fact that we need to transform the total
weights between groups of stations into weights for each individual station. As
shown in~\cite{wifi-slicing}, this is conceptually just a simple division.
However, when implementing this in an operating system kernel, we are limited to
integer arithmetic, which means that to get accurate weights, we need to ensure
that the division works when confined to the integers. To achieve this, we first
limit our configuration language to be expressed as integer weights between
groups. Then, to ensure that we can divide these weights with the number of
active stations, we multiply them by a suitable constant, chosen as follows:

We are given the set of groups $I$, where each group $i$ has a configured group
weight $W_i$ and $N_i$ active stations. We then define the multiplication
constant $C = \prod_{i\in I}N_i$. Multiplying all group weights by this same
constant maintains their relative ratio, and the choice of constant ensures that
each group's weight can be divided by the number of active stations in that
group. This gives us the following expression for the per-station weight for
group $i$:

\begin{equation}
  \label{eq:1}
  W^s_{i} = \frac{W_iC}{N_i}
\end{equation}

The third issue we need to deal with is converting the weights to the
per-station time quantums that are used in the scheduler, and which are
expressed in microseconds. These should be kept at a reasonable absolute size,
because larger weights result in coarser time granularity of the scheduler,
making each scheduler round take longer and impacting latency of all devices in
the network. We convert the calculated weights into final quantums by
normalising them so they fall within a range of $100-1000\, \mu s$, but
preferring smaller values if the ratio between the smallest and largest weight
is more than $10\times$.

\subsection{Kernel Airtime Scheduler}
\label{sec:kernel-scheduler}
We implement the kernel part of PoliFi in the WiFi protocol subsystem of the
Linux kernel (called \emph{mac80211}). Our implementation builds on our previous
airtime fairness scheduler, described in~\cite{usenix-paper}, which implemented
a queueing system in this layer. In this queueing system, packets are assigned a
Traffic ID (TID) before enqueue, and a separate queueing structure is created
for each TID, of which there are 16 per station. These per-TID queues then form
the basis of the scheduling of different stations. The queueing structure itself
is based on the FQ-CoDel queue management scheme~\cite{rfc8290} and ensures flow
isolation and low queueing latency.

While our previous implementation implemented queueing in the general WiFi
layer, scheduling and tracking of each active station's airtime usage was still
the responsibility of the driver. In PoliFi, we move the scheduling decisions
into mac80211, where it can be leveraged by all device drivers. In addition, we
modify the scheduler to support the weight-based policy enforcement capability
described above. The weights can be set by userspace through the standard
\emph{nl80211} configuration API.

In order to move the scheduling decision out of the drivers, we define a new
driver API, shown in Algorithm~\ref{alg:airtime-scheduler}. The device driver
runs the \texttt{schedule()} function, and asks mac80211 for the next TID queue
to schedule, using the \texttt{get\_next\_tid()} API function. The driver then
services this queue until no more packets can be scheduled (typically because
the hardware queue is full, or the TID queue runs empty). After this, the driver
uses the \texttt{return\_tid()} API function to return the TID queue to the
scheduler. A third API function, \texttt{account\_airtime()}, allows the driver
to register airtime usage for each station, which is typically done
asynchronously as packets are completed or received.

\begin{algorithm}[tb]
\caption{\small Airtime fairness scheduler. The schedule function is part of the device driver and is called on packet arrival and on transmission completion. The account\_airtime function is called by the driver when it receives airtime usage information on TX completion or packet reception.}
\label{alg:airtime-scheduler}
\begin{algorithmic}[1]
\small
\Function {schedule}{\textit{qoslvl}}
    \State $\textit{tid} \gets$ \Call{get\_next\_tid}{\textit{qoslvl}}
    \State \Call{build\_aggregate}{\textit{tid}}
    \State \Call{return\_tid}{\textit{tid}}
\EndFunction
\Function {get\_next\_tid}{\textit{qoslvl}}
    \State \emph{tid} $\gets$ \Call{find\_first}{\emph{active\_tids}, \textit{qoslvl}}
    \State \emph{stn} $\gets \textit{tid.station}$
    \State $\textit{deficit} \gets \textit{stn.deficit[qoslvl]}$
    \If {$\textit{deficit} \leq 0$}
      \State $\small\textit{stn.deficit[qoslvl]} \gets $ {\small$\textit{deficit} + \textit{stn.weight}$ }
      \State \Call{list\_move\_tail}{\emph{tid}, \emph{active\_tids}}
      \State \textbf{restart}
    \EndIf
    \State \Call{list\_remove}{\emph{tid}, \emph{active\_tids}}
    \State \Return {\textit{tid}}
\EndFunction
\Function {return\_tid}{\textit{tid}}
    \If {\textit{tid.queue} is not empty}
      \State \Call{list\_add\_head}{\emph{tid}, \emph{active\_tids}}
    \EndIf
\EndFunction
\State ~
\Function {account\_airtime}{\textit{tid}, \textit{airtime}}
    \State \emph{stn} $\gets \textit{tid.station}$
    \State \emph{qoslvl} $\gets \textit{tid.qoslvl}$
    \State $\textit{stn.deficit[qoslvl]} \gets \textit{stn.deficit[qoslvl]} - \textit{airtime}$
\EndFunction
\end{algorithmic}
\end{algorithm}

Using this API, mac80211 has enough information to implement airtime policy
enforcement using the weighted deficit scheduler approach described above. As
for the previous airtime fairness scheduler in the driver, airtime deficits are
kept separately for each of the four hardware QoS levels, to match the split of
the hardware transmission queue scheduling. The algorithm is implemented as part
of the \texttt{get\_next\_tid()} function as shown in
Algorithm~\ref{alg:airtime-scheduler}. The PoliFi scheduler has been accepted
into the upstream Linux kernel and will appear in Linux 5.1 with support for the
\emph{ath9k} and \emph{ath10k} drivers for Qualcomm Atheros 802.11n and 802.11ac
hardware.

\section{Evaluation}
\label{sec:evaluation}
In this section we evaluate how effectively PoliFi is able to implement the
desired policies. We examine steady state behaviour as well as the reaction time
of the dynamic and limit modes with a changing number of active stations. To
show how airtime policies can provide benefits for specific applications, we
also include an DASH video streaming use case in our evaluations. We perform the
experiments on our testbed with four WiFi devices. The details of our setup are
omitted here due to space constraints, but are available in an online
appendix~\cite{polifi_zenodo}.

\begin{figure}
  \centering
  \begin{subfigure}{\linewidth}
    \includegraphics[width=\linewidth]{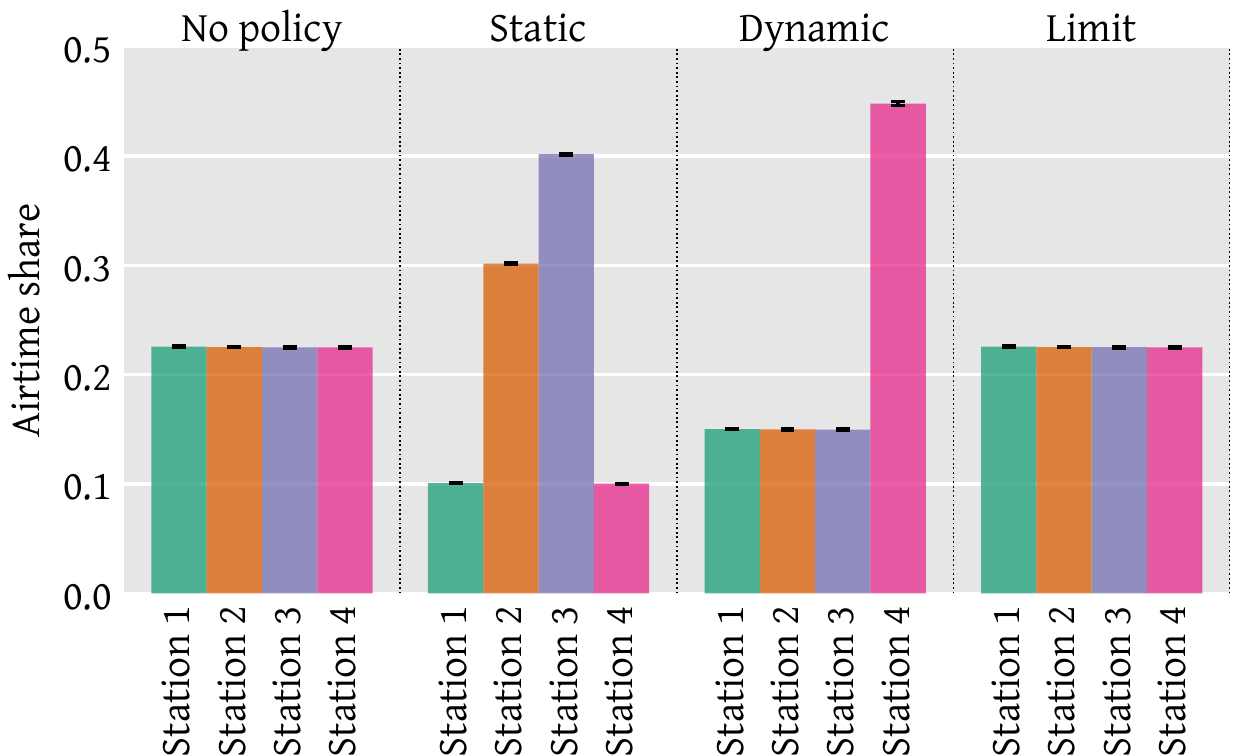}
    \caption{UDP traffic}
  \end{subfigure}

  \begin{subfigure}{\linewidth}
    \includegraphics[width=\linewidth]{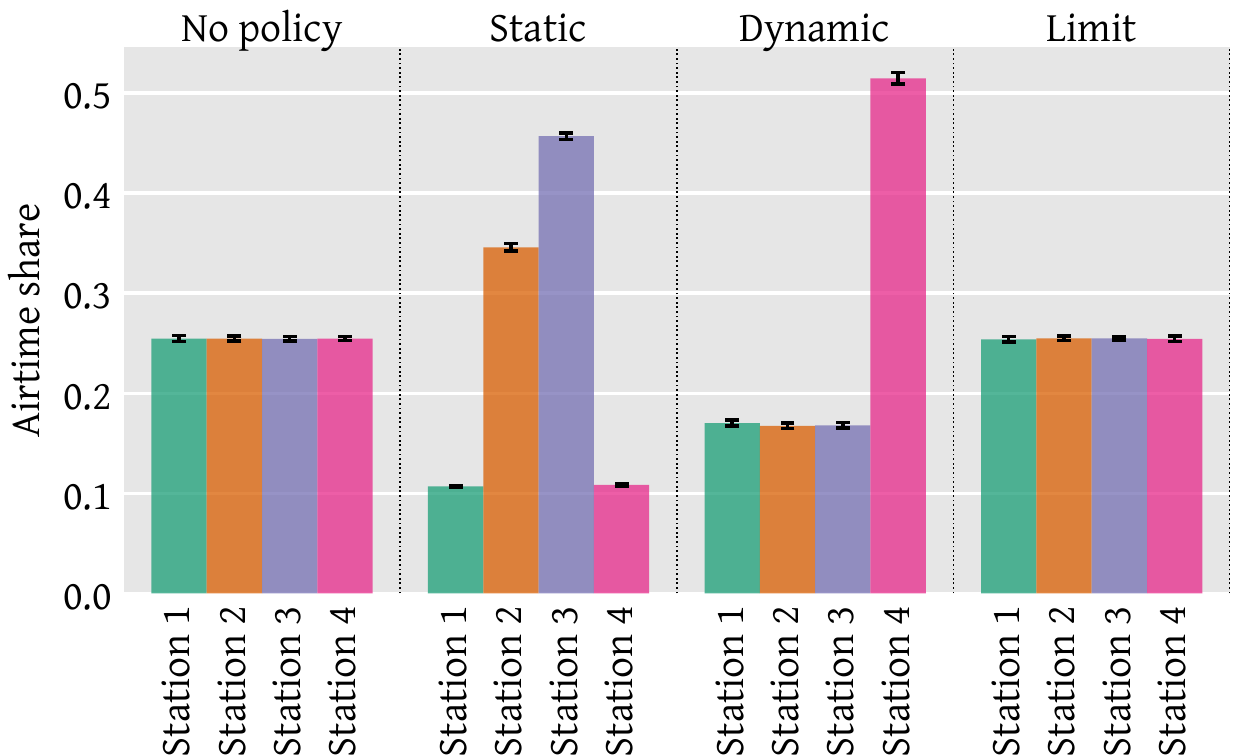}
    \caption{TCP traffic}
  \end{subfigure}
  \caption{\label{fig:steady-stations}Aggregate airtime usage share of four
    stations, over a 30-second bulk transfer. Graph columns correspond to the
    different policy modes. In static mode stations 2 and 3 are assigned weights
    of 3 and 4, respectively. In dynamic and limit mode, stations 1-3 are on one
    BSS while station 4 is on another; both BSSes have the same weight, and the
    second BSS is set to limited. The plots are box plots of 30 test runs, but
    look like lines due to the low variation between runs.}
\end{figure}

\begin{figure}
  \centering
  \begin{subfigure}{0.48\linewidth}
    \includegraphics[width=\linewidth]{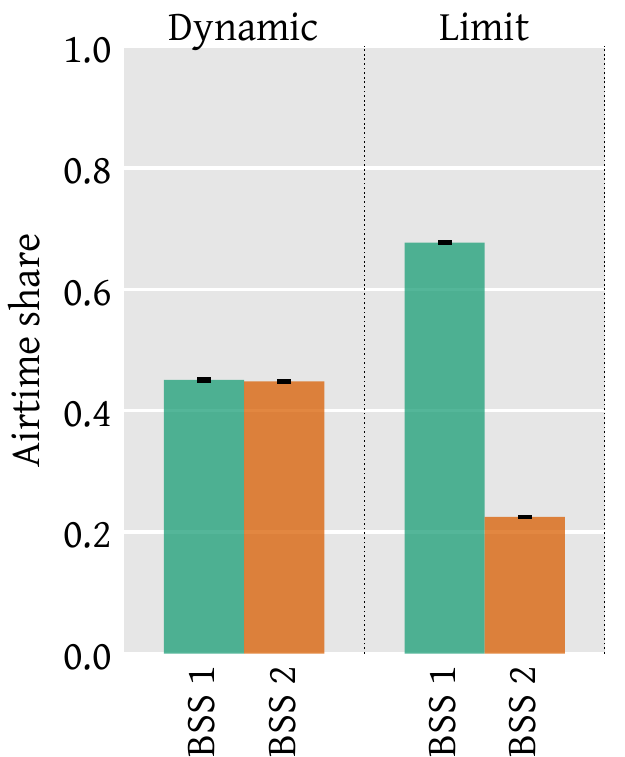}
    \caption{UDP traffic}
  \end{subfigure}
  ~
  \begin{subfigure}{0.48\linewidth}
    \includegraphics[width=\linewidth]{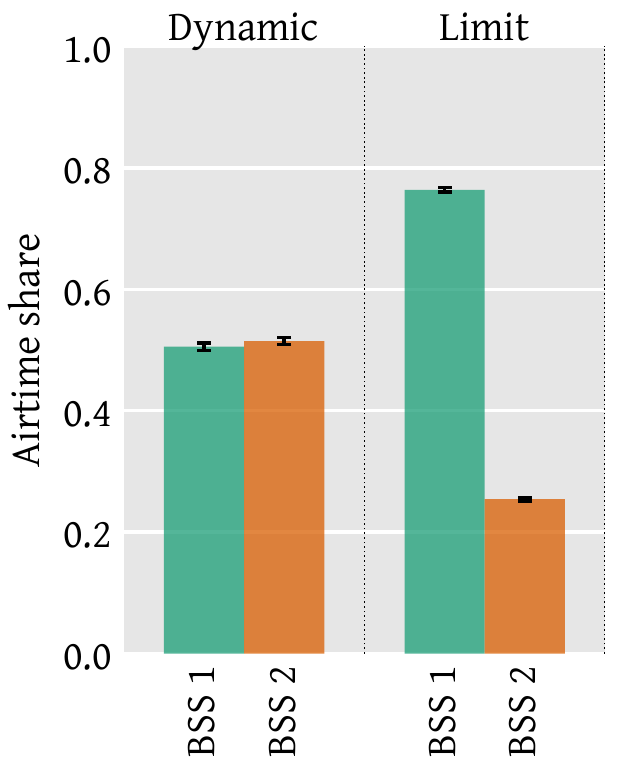}
    \caption{\label{fig:steady-bss-tcp}TCP traffic}
  \end{subfigure}
  \caption{\label{fig:steady-bsses}Aggregate airtime usage of the two BSSes, for
    the same test as that shown in Figure~\ref{fig:steady-stations}.}
\end{figure}

\subsection{Steady state measurements}
\label{sec:steady-state-meas}

The steady state tests consist of running a bulk flow (either UDP or TCP) to
each of four stations associated to the access point running PoliFi. Three of
the stations are associated to one BSS on the access point, while the fourth is
on a separate BSS. These two BSSes are the groups the algorithm balances in
dynamic and limit mode. Both groups are given equal weights, meaning that they
should receive the same total airtime share. When testing the limit mode use
case, the BSS with only a single station in it is set to \emph{limited}, which
in this case means that its natural airtime share is less than the configured
share, and thus that no limiting is necessary to enforce the configured policy.
We test this to ensure that the algorithm correctly allows the group that is not
marked as limited to exceed its configured airtime share.

The aggregate airtime usage of the stations and BSSes is seen in
figures~\ref{fig:steady-stations} and~\ref{fig:steady-bsses}, respectively. With
no policy configured, the scheduler simply enforces fairness between the active
stations. In the static policy mode, relative weights of 3 and 4 are assigned to
stations 2 and 3, respectively. These weights are clearly reflected in the
airtime shares achieved by each station in the second column of the graphs in
Figure~\ref{fig:steady-stations}, showing that static policy assignment works as
designed.

Turning to the group modes, Figure~\ref{fig:steady-bsses} shows the aggregate
airtime for each of the two configured BSSes. In dynamic mode, the scheduler
enforces equal sharing between the two BSSes, which translates to the single
station in BSS 2 getting three times as much airtime as the other three, as is
seen in the third column of Figure~\ref{fig:steady-stations}. In limit mode, BSS
2 is limited to at most half of the airtime, but because there is only one
station connected to it, its fair share is already less than the limit, and so
this corresponds to the case where no policy is enforced. Thus, the tests show
that the scheduler successfully enforces the configured policies for all three
use cases.

\subsection{Dynamic measurements}
\label{sec:dynamic-measurements}

\begin{figure}
  \centering
  \begin{subfigure}{\linewidth}
    \includegraphics[width=\linewidth]{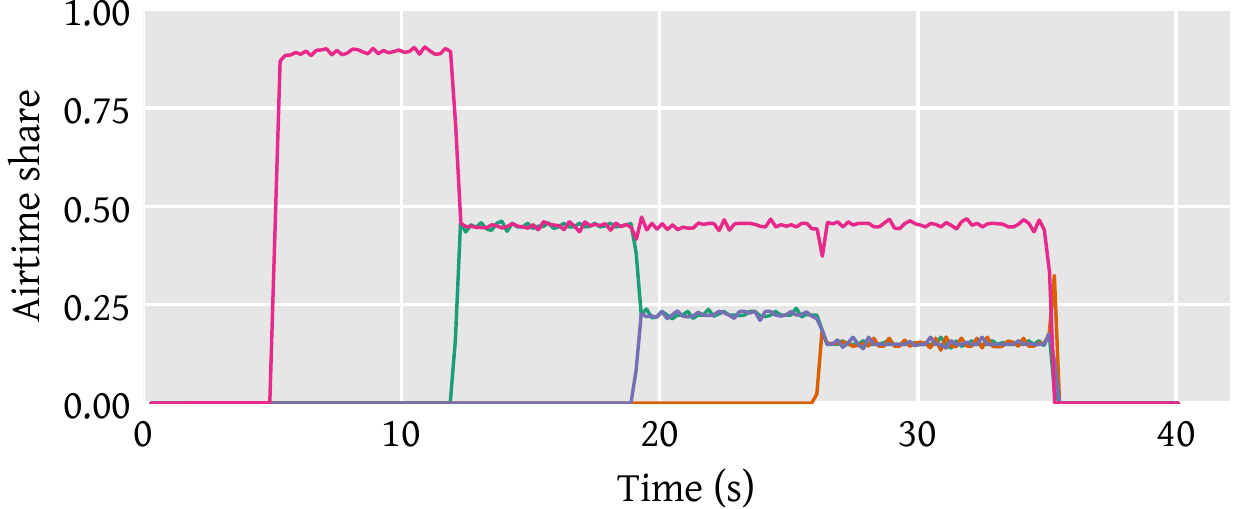}
    \caption{\label{fig:airtime-stagger-dynamic}Dynamic mode}
  \end{subfigure}
  \begin{subfigure}{\linewidth}
    \includegraphics[width=\linewidth]{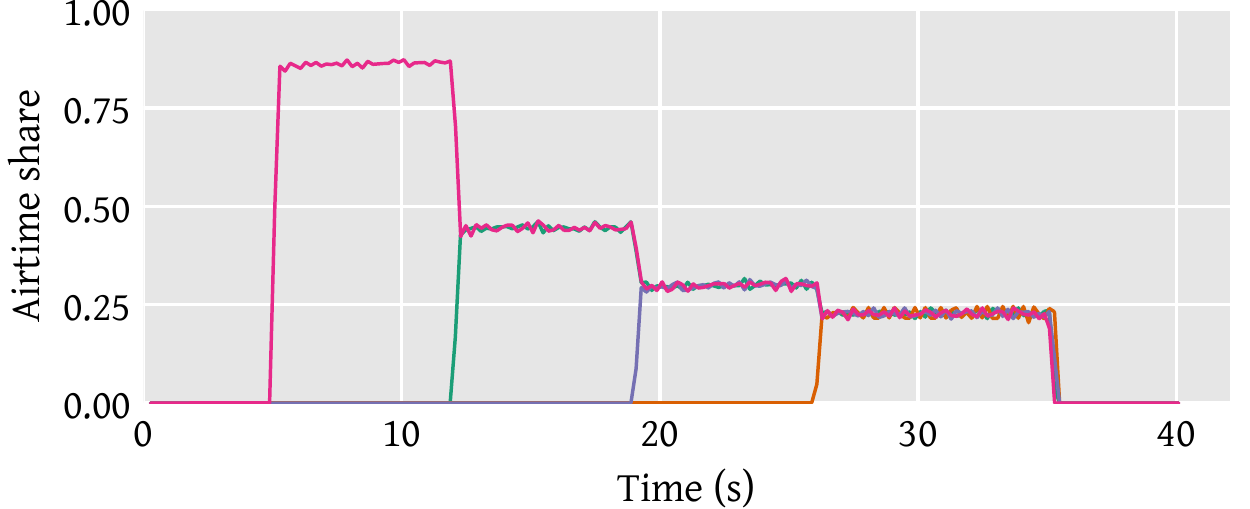}
    \caption{\label{fig:airtime-stagger-limit}Limit mode}
  \end{subfigure}
  \caption{\label{fig:airtime-stagger}Airtime usage over time with changing
    number of active stations, in dynamic and limit mode. UDP flows to each
    station start 5 seconds apart. The purple station (starting first) is on one
    BSS, while the remaining three stations are on the other BSS.}
\end{figure}

To evaluate the reaction time of the scheduler as station activity varies, we
perform another set of UDP tests where we start the flows to each of the
stations five seconds apart. We perform this test for the dynamic and limit
modes, as these are the cases where the scheduler needs to react to changes in
station activity.

The results of this dynamic test is shown in Figure~\ref{fig:airtime-stagger} as
time series graphs of airtime share in each 200\,ms measurement interval. The
station that starts first is Station 4 from the previous graphs, i.e., the
station that is on BSS 2. In dynamic mode, as seen in
Figure~\ref{fig:airtime-stagger-dynamic}, the first station is limited to half
the available airtime as soon as the second station starts transmitting. And
because the two groups are set to share the airtime evenly, as more stations are
added, the first station keeps using half the available airtime, while the
others share the remaining half.

In limit mode, as we saw before, the airtime shares of each of the four stations
correspond to their fair share. This is also seen in
Figure~\ref{fig:airtime-stagger-limit}, where all stations share the airtime
equally as new stations are added.

These dynamic results show that PoliFi has a short reaction time, and can
continuously enforce airtime usage policies as station activity changes. This is
important for deployment in a real network with varying activity levels.

\subsection{DASH Traffic Test}
\label{sec:dash-traffic-test}
To showcase an example real-world use case that can be improved by airtime
policy enforcement, we examine a DASH video streaming application. In this
scenario, we add a station with poor signal quality to the network, representing
a streaming device that is connected to the wireless network at a location where
signal quality is poor. Moving the device is not an option, so other measures
are necessary to improve the video quality. We stream the Big Buck
Bunny~\cite{bunny} video using the dash.js~\cite{dash-js} player running in the
Chromium browser on the slow station. We determine that the maximum video
bitrate the device can reliably achieve in this scenario (with no competing
traffic) is 2\,Mbps. However, when the other devices are active, the video
bitrate drops to 1\,Mbps because of contention.

\begin{figure}
  \centering
  \includegraphics[width=\linewidth]{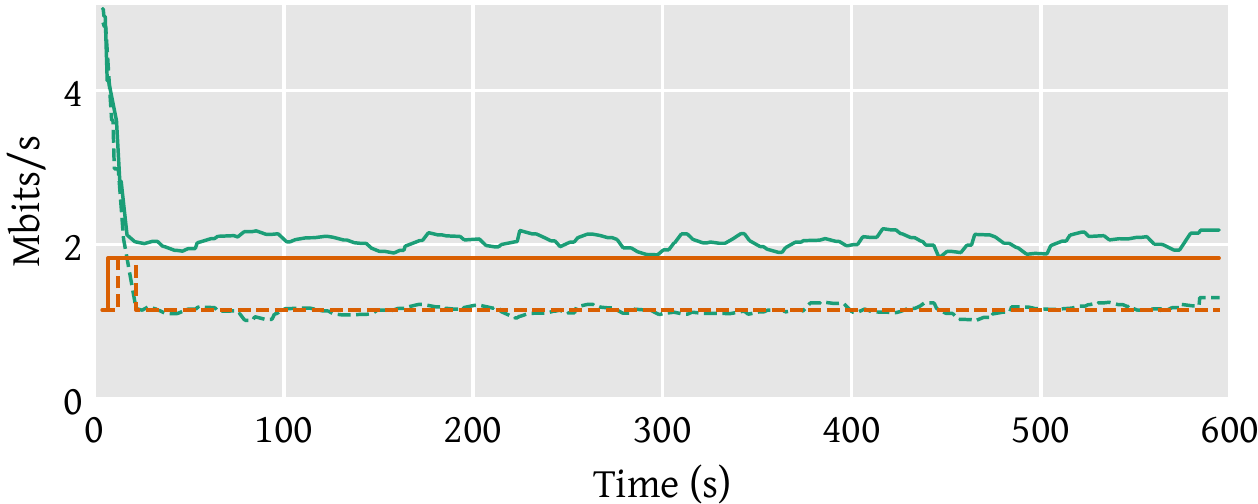}
  \caption{\label{fig:dash-throughput}DASH video throughput with prioritisation
    (solid lines) and without (dashed lines). The straight lines (orange) show
    the video bitrate picked by the player, while the others show the actual
    data stream goodput.}
\end{figure}

Figure~\ref{fig:dash-throughput} shows the achieved video bitrate along with the
data goodput of the video flow, while three other stations are simultaneously
receiving bulk data. With no policy set, the video bitrate drops to 1\,Mbps, as
described above. However, when we prioritise the station (to half the available
airtime in this case), the achieved bitrate stays at 2\,Mbps throughout the
10-minute video. This shows how PoliFi can improve the performance of a specific
real-world application.

\section{Conclusion}
\label{sec:conclusion}
We have presented PoliFi, a solution for enforcing airtime usage policies in
WiFi networks. Our evaluation shows that PoliFi makes it possible to express a
range of useful policies, including prioritisation of specific devices, and
balancing or limiting of groups of devices. We have also shown how the policy
enforcement can improve the performance of a real-world DASH video streaming
application.

PoliFi can improve performance of WiFi networks with high airtime contention,
and enables novel network usages such as network slicing. For this reason we
believe it to be an important addition to modern WiFi networks, which is made
widely available through its inclusion in the upstream Linux WiFi stack.

\bibliographystyle{IEEEtran}
\bibliography{phd,bufferbloat,rfc,wifi}
\end{document}